\documentclass[aps,prb,twocolumn,superscriptaddress]{revtex4-1}

\usepackage{graphicx}
\usepackage{bm}
\usepackage{amsmath}
\usepackage{amssymb}
\usepackage{hyperref}
\usepackage[version=4]{mhchem}
\usepackage{color}
\usepackage{upgreek}

\begin{document}

\title{Trapping molecules in a covalent graphene-nanotube hybrid}

\author{Zhao Wang}
\email{zw@gxu.edu.cn}
\affiliation{Laboratory for Relativistic Astrophysics, Department of Physics, Guangxi University, Nanning 530004, China}

\begin{abstract}

This study employs molecular dynamics simulations to examine the physisorption behavior of hydrocarbon molecules on a covalent graphene-nanotube hybrid nanostructure. The results indicate that the adsorbed molecules undergo self-diffusion into the nanotubes without the need for external driving forces, primarily driven by significant variations in binding energy throughout different regions. Notably, these molecules remain securely trapped within the tubes even at room temperature, thanks to a ``gate'' effect observed at the neck region, despite the presence of a concentration gradient that would typically hinder such trapping. This mechanism of passive mass transport and retention holds implications for the storage and separation of gas molecules.

\end{abstract}


\maketitle

\section{Introduction}

Reversible gas physisorption plays a crucial role in various applications like hydrogen storage, sensing, carbon capture, pollutant removal, and the separation and purification of chemical substances.\cite{Li2009} For these purposes, an ideal adsorbent should possess high adsorption capacity, selectivity, and favorable adsorption kinetics and regenerability. Carbon nanotubes (CNTs) and graphene have emerged as promising gas adsorption materials due to their large specific surface area, high chemical inertness, light mass density, outstanding mechanical strength, and thermal conductivity.\cite{Wang2013} However, each of these materials have their own advantages and disadvantages. Graphene, for example, is an atomic layer with an extreme theoretical specific surface area, yet its gas adsorption capacity is highly anisotropic due to its default inter-layer spacing of only $0.34$ nm. Conversely, CNT arrays are capable of adsorbing and storing molecules in three dimensions (3D), yet their narrow entrances can result in low adsorption efficiency.

Graphene-nanotube hybrids (GNHs) have been synthesized to offer the advantages of both graphene and CNTs,\cite{Dasgupta2016,Lv2014} with the potential of adsorbing gas molecules onto a graphene surface and transporting them into CNTs for 3D storage. A variety of GNH structures have been designed and characterized for molecular adsorption-related applications.\cite{Dimitrakakis2008,Wesolowski2011,Wu2012,Lei2014,Wesolowski2016,Zhou2016} However, while previous studies have considered external pressure or pre-positioning of molecules on the GNH surface, a solution for driving the motion of molecules within the GNH has not been proposed.

The literature describes several active methods for transporting molecules or particles on the surface of nanomaterials, including the application of a thermal gradient,\cite{Schoen2006,Barreiro2008,Wang2019,Wang2020} electric,\cite{Ohara2008,Dong2009} and mechanical means.\cite{Russell2012,Wang2007,Meng2020,Wang2019a} However, implementing these active techniques on nanostructures is very challenging. Heuristically, theoretical calculations have revealed that the binding energy of molecules inside CNTs is higher than that on a graphene layer, implying that molecules in CNTs could be more stable than those on graphene.\cite{Henwood2007,Fan2009,Denis2010,Kondratyuk2005} Inspired by this observation, we conducted molecular dynamics (MD) to simulate the physisorption of hydrocarbon molecules on a GNH nanostructure. The results indicated that, in a GNH, the molecules could be efficiently transported into CNTs at room temperature, and then firmly trapped inside the CNTs until released by heating. These findings provide a solution for the passive transportation and retention of molecules within GNHs, eliminating the need for an external driving force.
 
\section{Methods}

\begin{figure}[htp]
\centerline{\includegraphics[width=9cm]{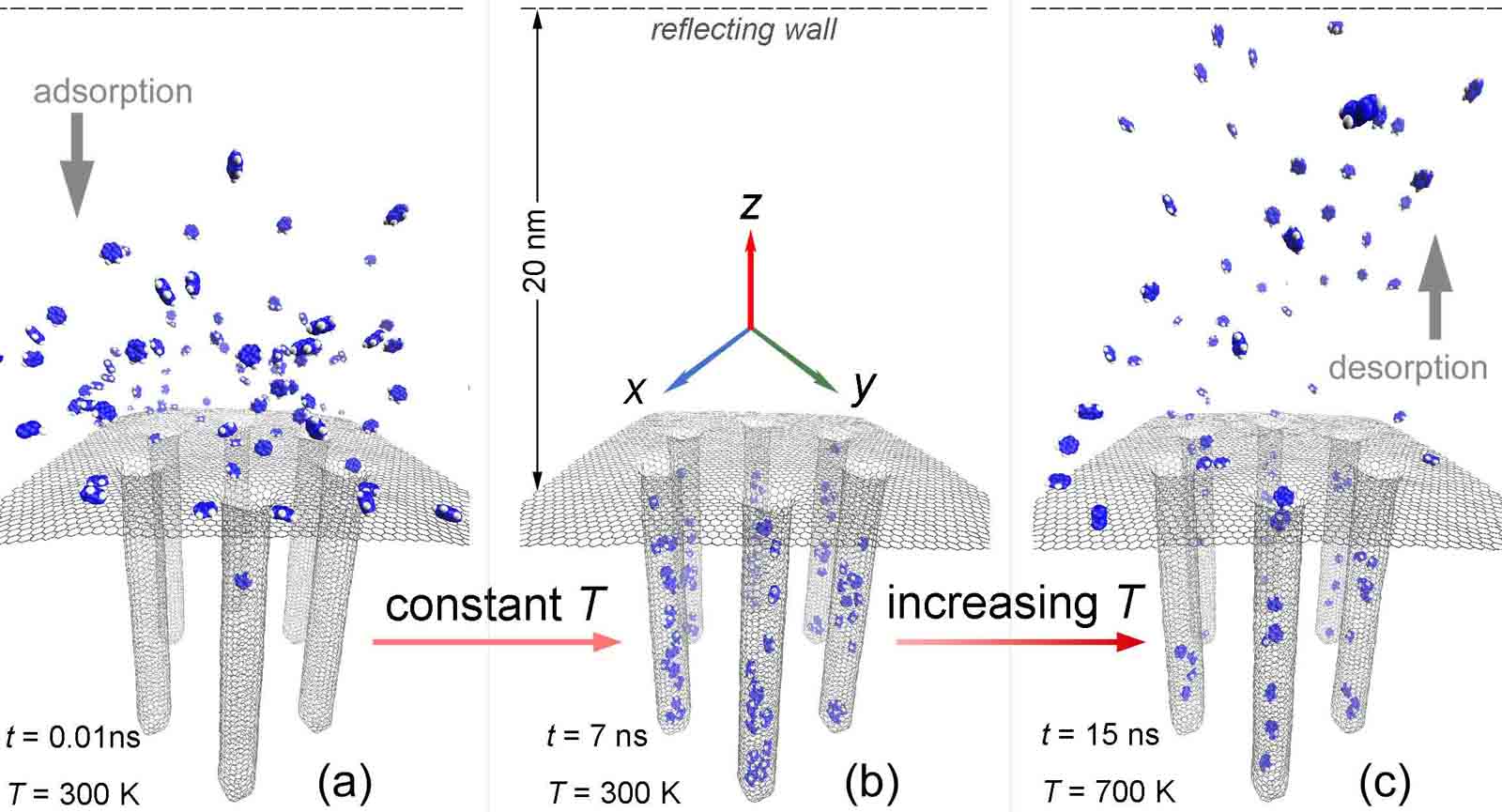}}
\caption{\label{F1}
Snapshots of the simulation cell at different time $t$ and temperature $T$: (a) $t=0.01$ ns and $T=300$ K, (b) $t=7$ ns and $T=300$ K, (c) $t=15$ ns and $T=700$ K.}
\end{figure}

The GNH nanostructure was composed of six (10,10) CNTs connected to a layer of graphene via cylindrical holes. As illustrated in Figure \ref{F1}(a), the tubes were $11$ nm in length and $1.38$ nm in diameter, with their opposing ends capped by the hemispheres of C240. The tubes were axially rotated to fit the graphene holes, resulting in the formation of heptagon and hexagon rings at their junctions, thus creating a stable configuration.\cite{Xu2012} The simulation cell was periodic in the graphene plane, with an area of $x \times y = 6.42 \times 6.30$ nm$^{2}$. The 3rd dimension was $20$ and $13$ nm along $+z$ and $-z$ from the graphene plane, respectively, resulting in a free space of approximately $905$ nm$^{3}$, including the volume of the tubes. 

The simulation was performed in three stages. Initially, benzene molecules were randomly placed in the upper space, as shown in Figure \ref{F1}(a). The system was run at $T_{1}=300$ K for $7.5$ ns in a canonical ensemble with a Nos\'{e}-Hoover thermostat applied to the GNH, resulting in configurations e.g. that in Figure \ref{F1}(b). Subsequently, the temperature was linearly increased to a higher temperature $T_{2}$ over $3$ ns, and was kept constant at $T_{2}$ for $9.5$ ns in order to release the adsorbates, as shown in Figure \ref{F1}(c). The MD was run for $20$ ns in $4 \times 10^{7}$ steps at a time step of $0.5 \times 10^{-6}$ ns, using the Verlet algorithm for time integration as implemented in the Large-scale Atomic/Molecular Massively Parallel Simulator (LAMMPS).\cite{Plimpton1995}

The Adaptive Interatomic Reactive Empirical Bond Order (AIREBO) force field was used to describe interatomic interactions, which incorporate both covalent and non-covalent contributions. The covalent component is described by the Reactive Empirical Bond Order (REBO) model,\cite{Brenner2002} which considers the influence of bond length, bond angle, dihedral angle, bond order, and torsion of single bond. Meanwhile, the non-covalent component is described by a $6-12$ Lennard-Jones term to account for long-range van der Waals (vdW) interactions. Spline regression is applied to provide a smooth transition between covalent and non-covalent interactions, with cutoff radii of $0.2$ and $1.1\;\mathrm{nm}$, respectively. AIREBO is widely used for hydrocarbon systems due to its high accuracy, despite of the expense of computation time. Its ability to describe bond rotation and torsion in terms of bond order is particularly important in the modeling of the adsorption process,\cite{Gao2021,Li2014} as the deformation of the adsorbate molecule induced by the substrate, or vice versa, can be accurately represented.\cite{Petucci2013,Hantal2010,Kostov2002,Ni2002,Hanine2020,Qi2018,Zhou2021,Raghavan2017}

\section{Results and Discussion}

\begin{figure}[htp]
\centerline{\includegraphics[width=9cm]{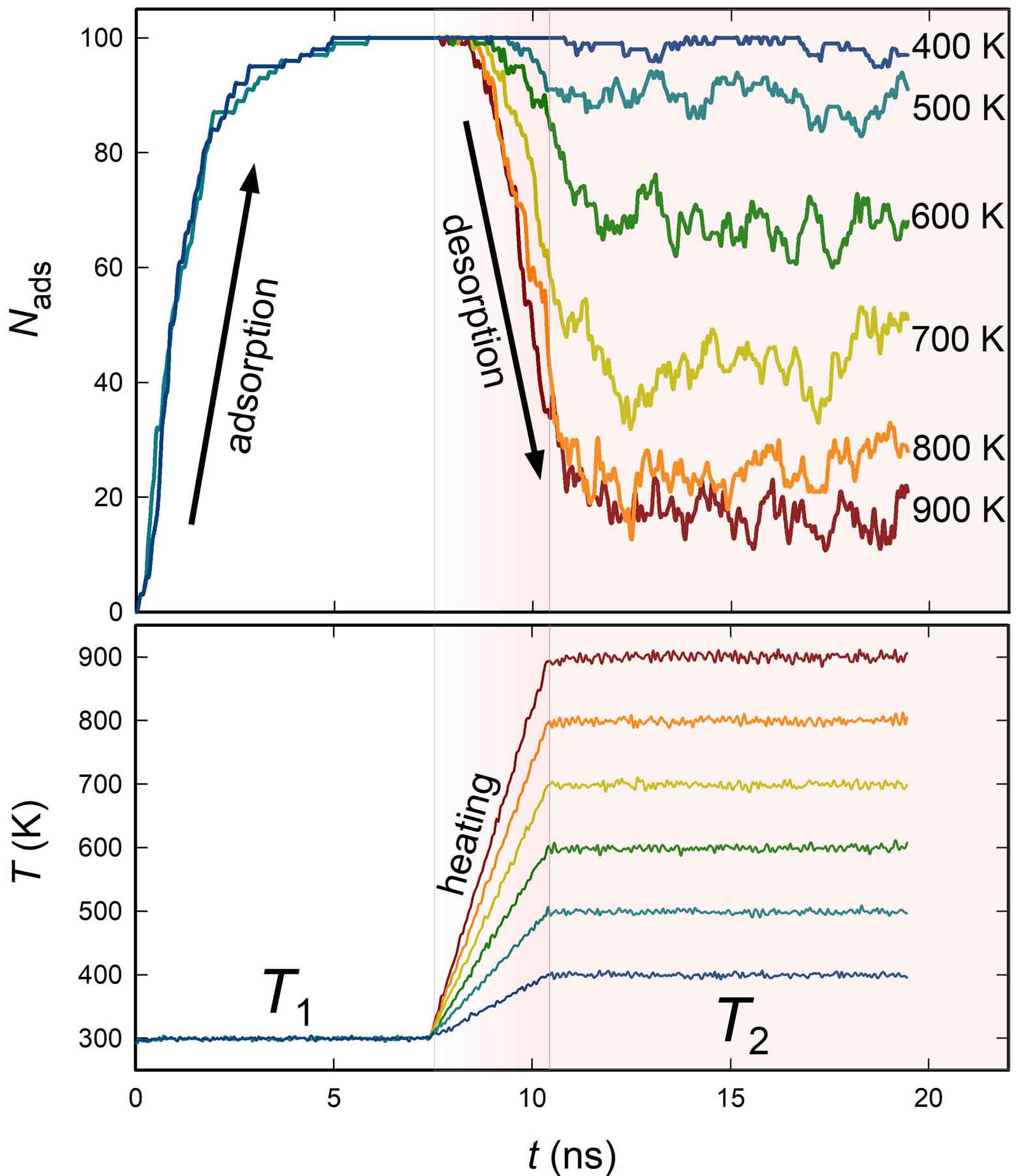}}
\caption{\label{F2}
Time evolution of the number of molecules (among 100) inside the CNTs (upper panel), and the corresponding system temperature (lower panel).}
\end{figure}

A set of simulations were conducted for $100$ benzene molecules on the GNH. At $T_{1}= 300$ K, the molecules were adsorbed, and then released at different temperatures ranging from $400$ to $900$ K. Figure \ref{F2} illustrates the evolution of the number of adsorbed molecules inside the CNTs. It is seen that the molecules were exhaustively captured and stored within the first $6$ ns at $300$ K. Subsequently, with increasing temperature, the adsorbates start to dissociate from the GNH. The number of released molecules depends on the target temperature $T_{2}$. For instance, more than $80$ molecules dissociated at $900$ K, while only less than $5$ did so at $400$ K. The dynamic process of molecules being adsorbed on the graphene surface and spontaneously moving into the tubes via self-diffusion is demonstrated in a video provided in Supporting Information, which shows the GNH working as a vacuum cleaner at nanoscale.

\begin{figure}[htp]
\centerline{\includegraphics[width=9cm]{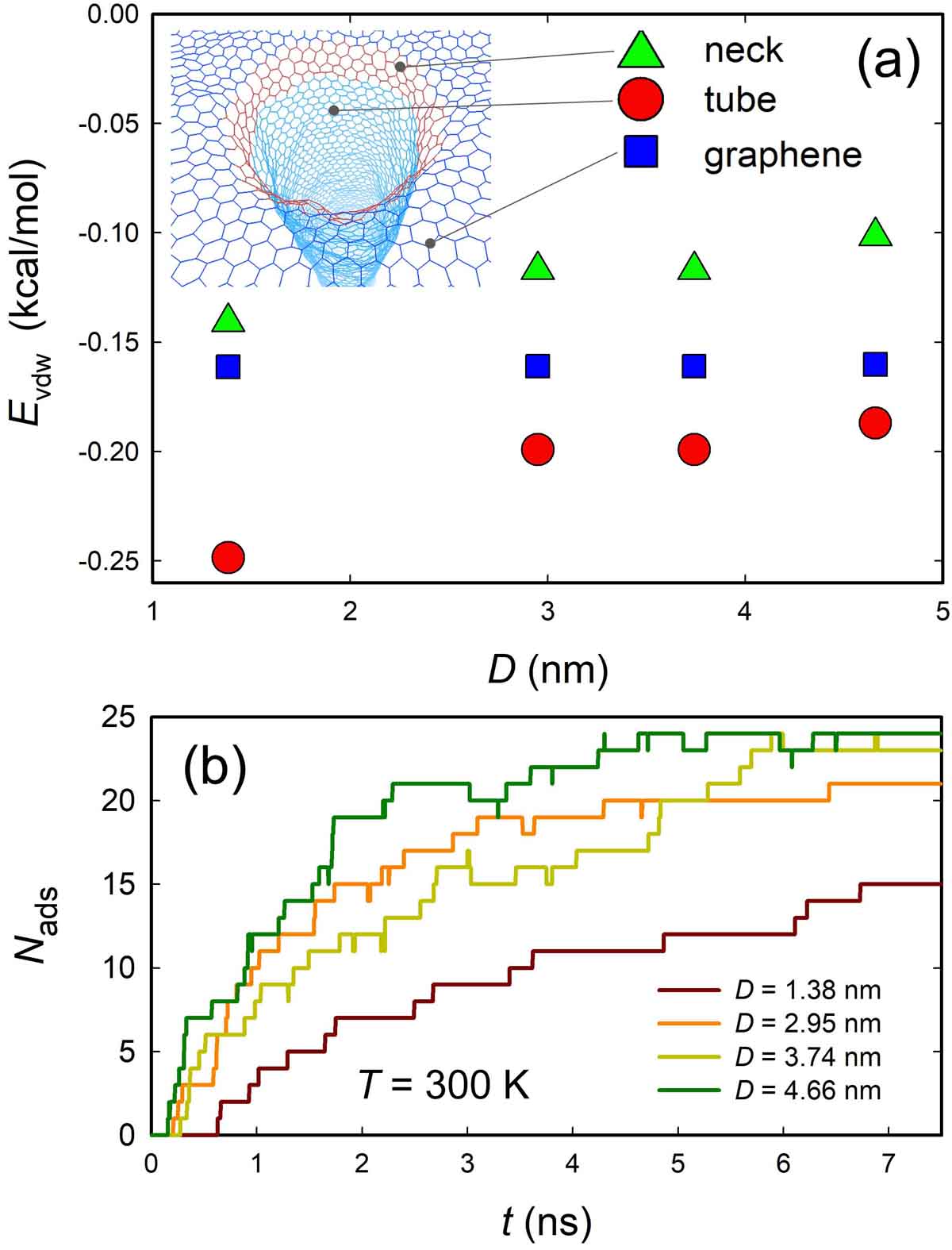}}
\caption{\label{F3}
(a) Mean binding energy (per molecule) between the GNH and the adsorbed benzene molecules on different sites vs the tube diameter, at a constant temperature of $300$ K. (b) Number of the molecules inside the CNT vs time for different tube diameters $D$.}
\end{figure}

Interestingly, the molecules firmly stayed inside the CNTs once they moved in during adsorption. This contradicts chemical intuition as the presence of a concentration gradient would typically hinder retention. To understand this, we conducted a set of simulations to calculate the vdW binding energy ($E_{vdw}$) of the adsorbate on different sites of the GNH. The simulations were performed at $300$ K with $25$ benzene molecules on a GNH composed of a single CNT with varying sizes. The average $E_{vdw}$ of the molecules in the CNTs was found to be almost twice higher than that of the ones on the neck regions, and illustrated in Figure\ref{F3} (a). These results indicate that the molecules may be more strongly bound to the CNTs than the neck regions, which explains their persistent presence inside the CNTs.

The variation in binding energy across the GNH can be attributed to the alterations in its crystalline structure. Specifically, the presence of 5-7-7-5 defects in the neck regions and the curvature of the tube surface contribute to the observed differences in binding energy. As depicted in Figure 3 (a), the binding energy disparity between graphene and the neck region falls within the range of $0.025-0.06$ kcal/mol. The variation in binding energy between the tube and graphene is in the range of $0.03-0.09$ kcal/mol, dependent upon the tube size. The molecules exhibit a preference for staying within the tubes because the binding energy is lower, rendering them more stable compared to the neck region. However, it is unlikely that we would observe the same effect if the outer surface is utilized for the same purpose, as the binding energies for molecules inside the CNTs are greater than those for adsorption outside the tubes, as suggested by Henwood and Carey.\cite{Henwood2007}

The binding energy for benzene molecules inside and on the surface of a CNT was found to be significantly lower than that of the molecules on the graphene. This is in agreement with previous calculations.\cite{Henwood2007,Fan2009,Denis2010} Interestingly, $E_{vdw}$ on the neck region was observed to be higher than on the other two regions. This induces a ``gate'' effect of the neck that locks molecules inside the CNT. Using the GNH with the CNT diameter of $3.74$ nm as an example, a molecule would need a kinetic energy of $\approx 0.4$ kcal/mol to overcome the barrier and move from the graphene to the CNT. However, it would require more than $0.8$ kcal/mol to move back to the graphene. This barrier difference could be a reason why molecules remain steady once they have moved into the CNTs.

\begin{figure}[htp]
\centerline{\includegraphics[width=9cm]{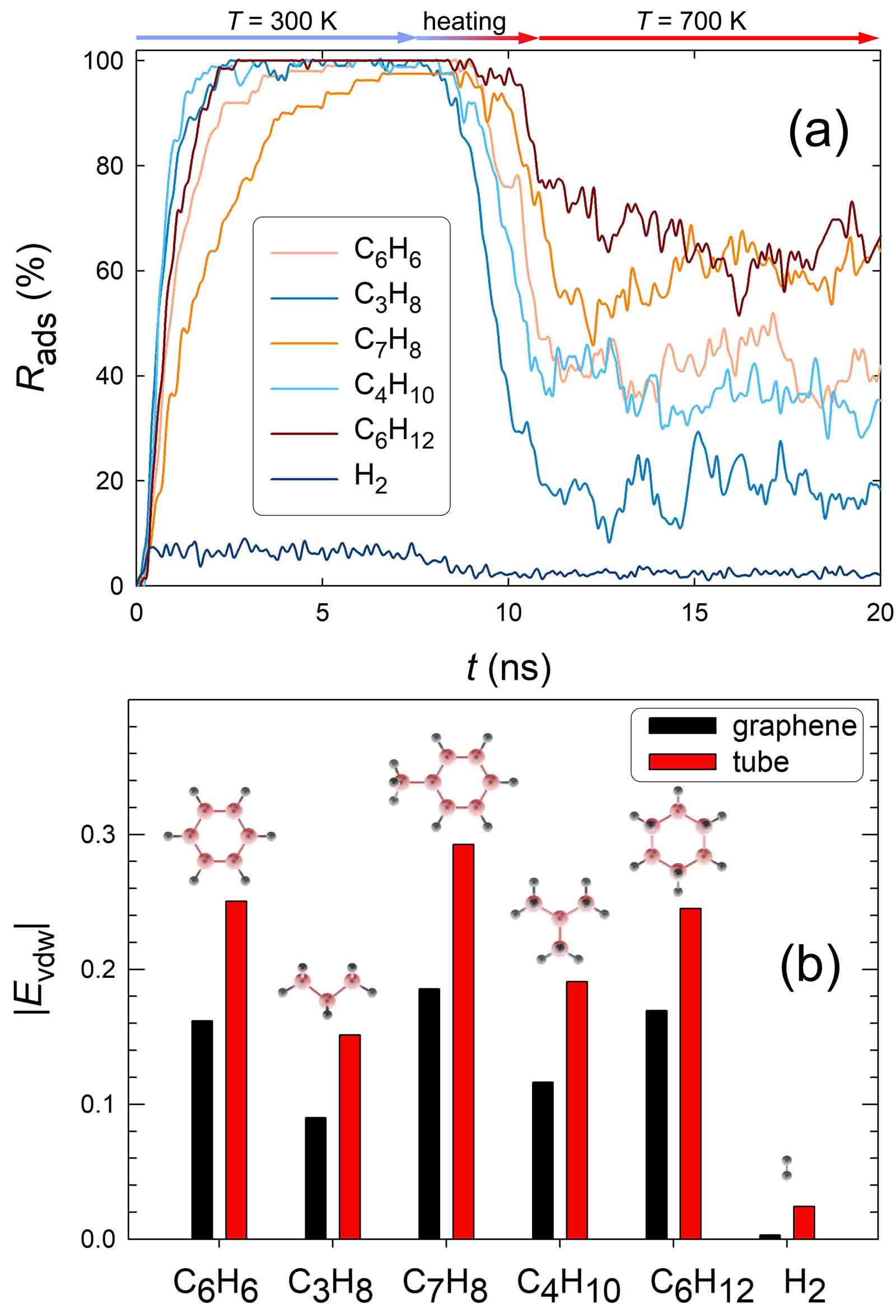}}
\caption{\label{F4}
(a) Ratio of the molecules in the CNTs vs time. (b) Absolute binding energy of molecules of different species on graphene (black bars) and in CNTs (red bars).}
\end{figure}

In addition to the gate effect, the collision of molecules with the neck region could also be inhibited within the CNT. e.g. the molecules are able to move freely to the neck region from all directions on graphene, whereas only those molecules that are close to the neck region have the potential to do so when they are in the CNT. The combination of the increased energy barrier at the neck region and the reduced collision probability due to the tube-structure may explain why molecules are mostly trapped within the CNT during the adsorption process. Furthermore, the simulation results show a correlation between the tube size and the speed of adsorption, as illustrated in Figure\ref{F3} (b). The largest tubes, with $D = 4.66$ nm, are able to adsorb molecules the quickest, whereas the smallest tube, with $D = 1.38$ nm, is seemingly filled with $15$ molecules.

We have simulated the adsorption and desorption processes of propane \ce{C3H8}, toluene \ce{C7H8}, isobutane \ce{C4H10}, cyclohexane \ce{C6H12} and \ce{H2}, in comparison to benzene on a GNH, to explore the potential of the ``locking'' mechanism. The GNH was composed of six CNTs as in Figure \ref{F1}. The simulations were conducted with a roughly constant gas concentration of 1.326 atom/nm$^{3}$, and the temperature was maintained at $T_{1}=300$ K for $7.5$ ns, before being heated up to $T_{2}=700$ K in $3$ ns, after which it remained constant at $T_{2}$ for another $9.5$ ns. The ratio of the molecules inside the CNTs with respect to their total numbers was calculated and is plotted in Figure\ref{F4} (a). The results demonstrate that the adsorption and desorption rates vary significantly among the different species, with \ce{C4H10} and \ce{C3H8} having the highest speeds, and \ce{C7H8}, \ce{C6H6}, and \ce{C6H12} adsorbing and dissociating more slowly. The observed differences in adsorption and desorption rates have potential implications for gas separation applications, leveraging the selectivity and binding energy variations.

The adsorption and desorption rates are inversely related to the absolute binding energy between the substrate and the adsorbates. Figure \ref{F4} (b) demonstrates that, due to their small molecular mass and aliphatic structure, the potential well depths of \ce{C4H10} and \ce{C3H8} are lower than those of other hydrocarbons. This results in faster adsorption and desorption of these molecules onto and from the GNH. Conversely, \ce{C7H8}, \ce{C6H12} and \ce{C6H6}, which possess higher molecular mass and aromatic-type geometry, experience strong vdW interactions that impede their removal from the GNH at high temperatures.

\begin{figure}[htbp]
\centerline{\includegraphics[width=9cm]{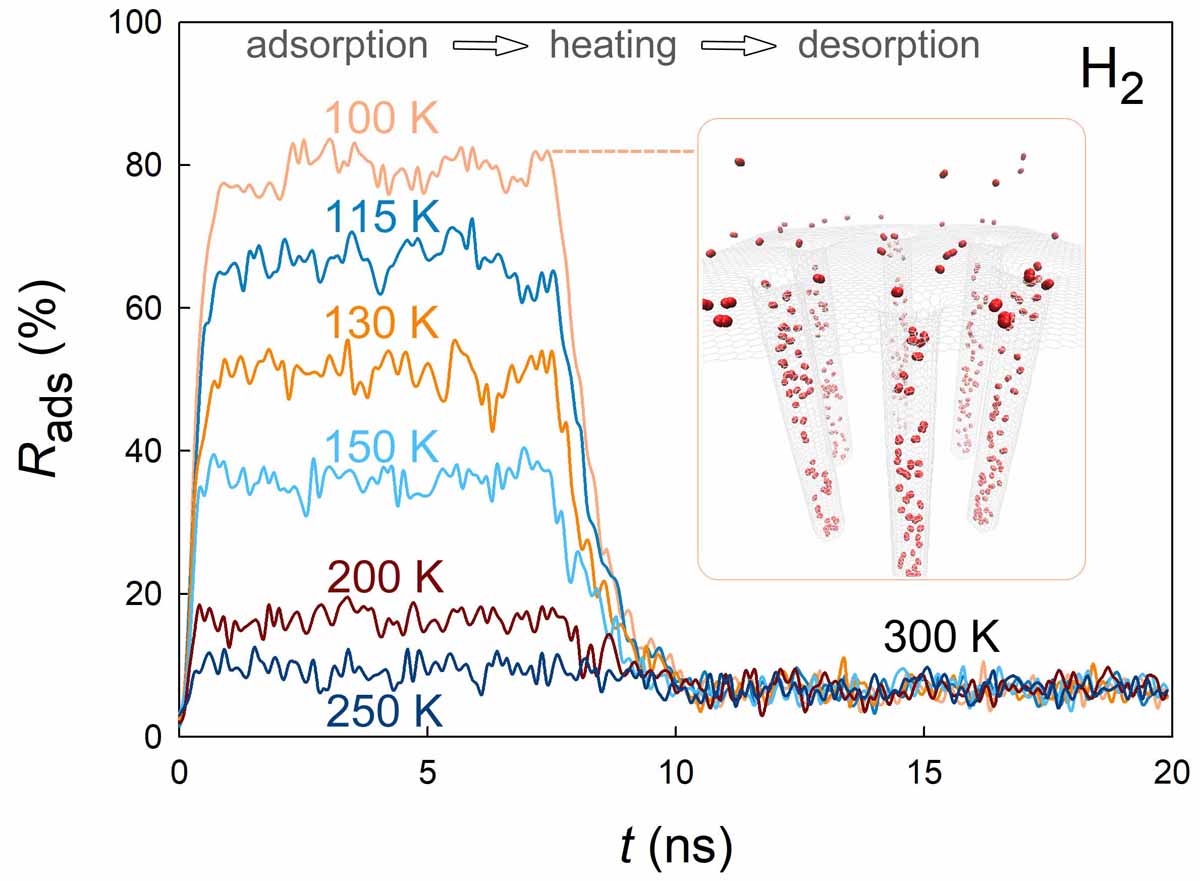}}
\caption{\label{F5}
Ratio of \ce{H2} molecules in the CNTs vs time at different temperatures. Inset: A snapshot of captured \ce{H2} at $100$ K when $t=7.5$ ns.}
\end{figure}

Simulations have demonstrated that \ce{H2} capture is sensitive to the temperature of the substrate. At $300$ K, only about $7\%$ of the total \ce{H2} mass was adsorbed into the CNTs. Further simulations have been performed for \ce{H2} capture, considering $T_{1}=100, 115, 130, 150, 200$, and $250$ K and $T_{2}=300$ K. Figure \ref{F5} shows that, when $T_{1}$ was set to $100$ K, approximately $80$\% of \ce{H2} was captured within the CNTs. This ratio decreased sensitively with increasing $T_{1}$, with only $20$\% captured at $200$ K and $10$\% at $250$ K. Thus, in order to achieve effective \ce{H2} capture, say 50\%, $T_{1}$ should be set to a temperature below $130$ K. We finally note that the temperature mentioned is currently unsuitable for practical use and requires improvement through chemical doping, as previously suggested by Dimitrakakis et al. \cite{Dimitrakakis2008}.

\section{Conclusions}
In this study, MD simulations demonstrated that the covalent GNH nanostructure has the ability to transport adsorbed molecules into CNTs without any external force, which remained inside at room temperature despite the concentration gradient. This intriguing gate effect arises from the distinct binding energy variations across the GNH, attributed to the presence of topological defects in the neck regions and the curvature of the tube surface. This discovery could potentially have implications for gas molecule storage and separation in covalent GNH nanostructures. The simulations conducted on different species also highlighted the distinct adsorption and desorption speeds of aromatic and aliphatic molecules, giving insight into the selective adsorption or separation of organic species using GNHs. Additionally, the study observed the gate effect in the adsorption of \ce{H2} despite low temperature.


\providecommand{\latin}[1]{#1}
\makeatletter
\providecommand{\doi}
  {\begingroup\let\do\@makeother\dospecials
  \catcode`\{=1 \catcode`\}=2 \doi@aux}
\providecommand{\doi@aux}[1]{\endgroup\texttt{#1}}
\makeatother
\providecommand*\mcitethebibliography{\thebibliography}
\csname @ifundefined\endcsname{endmcitethebibliography}
  {\let\endmcitethebibliography\endthebibliography}{}

\end{document}